\documentclass[conference]{IEEEtran}
\IEEEoverridecommandlockouts

\usepackage{graphicx,color,url}
\graphicspath{{./fig/}}
\usepackage{microtype}
\usepackage[all]{xy}
\usepackage{xspace}

\title{Software Defined Radio Architecture Survey for Cognitive Testbeds}

\author{\IEEEauthorblockN{Micka\"el Dardaillon, Kevin Marquet , Tanguy Risset, Antoine Scherrer}
\IEEEauthorblockA{
Universit\'{e} de Lyon, Inria,\\
INSA-Lyon, CITI, \\F-69621, Villeurbanne, France \\
Emails : \{mickael.dardaillon, kevin.marquet, tanguy.risset\}@insa-lyon.fr, antoine.scherrer@inria.fr
\thanks{This work is partially supported by R\'egion Rh\^one Alpes ADR 11 01302401.}
}}

\IEEEpubid{\makebox[\columnwidth]{978-1-4577-1379-8/12/\$26.00~\copyright~2012~IEEE} \hspace{\columnsep}\makebox[\columnwidth]{ }}

\begin{document}

\newcommand{\sdr}{\textsc{sdr}\xspace}

\newcommand{\asic}{\textsc{asic}\xspace}

\newcommand{\fpga}{\textsc{fpga}\xspace}

\newcommand{\cpu}{\textsc{cpu}\xspace}

\newcommand{\gpu}{\textsc{gpu}\xspace}

\newcommand{\gnu}{\textsc{gnu}\xspace}

\newcommand{\vhdl}{\textsc{vhdl}\xspace}

\newcommand{\adc}{\textsc{adc}\xspace}

\newcommand{\dac}{\textsc{dac}\xspace}

\newcommand{\ansic}{\textsc{ansi C}\xspace}

\newcommand{\arm}{\textsc{arm}\xspace}

\newcommand{\dsp}{\textsc{dsp}\xspace}

\newcommand{\gpp}{\textsc{gpp}\xspace}

\newcommand{\hal}{\textsc{hal}\xspace}

\newcommand{\imec}{Imec\xspace}

\newcommand{\lte}{\textsc{lte}\xspace}

\newcommand{\fit}{\textsc{fit}\xspace}

\newcommand{\mimo}{\textsc{mimo}\xspace}

\newcommand{\simd}{\textsc{simd}\xspace}

\newcommand{\vliw}{\textsc{vliw}\xspace}

\newcommand{\wcdma}{\textsc{wcdma}\xspace}

\newcommand{\risc}{\textsc{risc}\xspace}

\newcommand{\wimax}{\textsc{w}i\textsc{max}\xspace}

\newcommand{\ofdm}{\textsc{ofdm}\xspace}

\newcommand{\dma}{\textsc{dma}\xspace}

\newcommand{\fft}{\textsc{fft}\xspace}

\newcommand{\picoga}{\textsc{p}i\textsc{c}o\textsc{ga}\xspace}

\newcommand{\wifi}{\textsc{w}i\textsc{f}i\xspace}

\newcommand{\cortexlab}{{\em CortexLab}\xspace}

\maketitle

\begin{abstract}
In this paper we present a survey of existing prototypes dedicated to
software defined radio. We propose a classification related to the
architectural organization of the prototypes and provide some
conclusions about the most promising architectures. This study should
be useful for cognitive radio testbed designers who have to choose between
many possible computing platforms. We also
introduce a new cognitive radio testbed currently under
construction and explain how this study have influenced the test-bed
designers choices.
\end{abstract}


\section{Introduction}

Radio technologies have been developed in a static paradigm:
protocols, radio resources allocation and access network architecture
were defined beforehand, providing non-adaptable radio
systems. Nowadays, the saturation of radio frequency bands calls new
era of radio networking which will be characterized by self-adaptive
mechanisms. These mechanisms will rely on {\em software radio
  technologies}.

 The concept of  software radio has been coined by J. Mitola in
 his seminal work during the early 90's \cite{Mitola1992}.  While
 implementing the whole radio  node in software is still an utopia, many
 architectures now hitting the market include some degree of
 programmability.
%
Unfortunately, there is no agreement on
the hardware architecture embedded in a mobile terminal with \sdr
facilities. Various technologies are used: \asic, \fpga, \dsp,
\gpp, etc. These technologies are often mixed and sometimes the term
{configurable} is more adequate than programmable for them. 

Studying architectures for emerging {\sdr} systems is of crucial
importance because of the need to define the hardware abstraction
layer of {\sdr} systems: the {\em radio hardware abstraction layer}
({\sc r-hal}). We believe that it is also of first importance for
testbed providers, as we explain later in the paper.

In 2010, two important surveys where published~\cite{CatthoorSurvey,Ulversoy2010}.
In~\cite{Ulversoy2010}, Tore Ulvers{\o}y provides a very complete review of {\sdr} challenges
related to software architecture, computational requirements,
security, certification and business for {\sdr} systems. Some {\sdr}
architecture prototypes are mentioned but are not the main topic
of the study, and many other prototypes have been delivered since
2010. In~\cite{CatthoorSurvey}, Palkovic et al. provide a precise
comparative study between  the \imec Bear Platform and other
important \sdr multi-core architectures. The comparison is made
for architectures and programming flows. Our study focuses more
precisely on architectures of \sdr platforms available up to now.
Programming models and programming tools used in
these platforms are very important topics but much less mature today, most of
these platforms being currently programmed ``by hand''. All the performances results presented in this paper are taken from bibliography.

Another goal of our contribution is to show how such a technological
survey can be useful for cognitive radio testbed providers. In 2009, an
{\sc nsf} report~\cite{NSF_CR} pointed out the lack of cognitive radio testbeds and
urges {\em ``The development of a set of cognitive networking
  testbeds that can be used to evaluate cognitive networks at various
  stages of their development''}. We are currently involved
in the development of a cognitive radio testbed called
\cortexlab which is part of the \fit
platform~\cite{FIT}. Our work is motivated by the
development of this platform: what is the most adapted {\sdr} node
for a cognitive testbed accessed via Internet ?

The rest of the paper is organized as follows: section~\ref{techno} provides a brief summary of radio and {\sdr} technology.
Section~\ref{sec:platforms} describes the different platforms and gather them in categories.
The analysis of the different categories is made in the section~\ref{sec:analysis}.
Section~\ref{sec:FIT} describes the choices made for the {\em CortexLab} testbed.
We draw conclusions in section~\ref{sec:conclusion}.
 
\section{SDR technology}
\label{techno}

\begin{figure}[h]
	\centerline{
		\includegraphics[width=8cm]{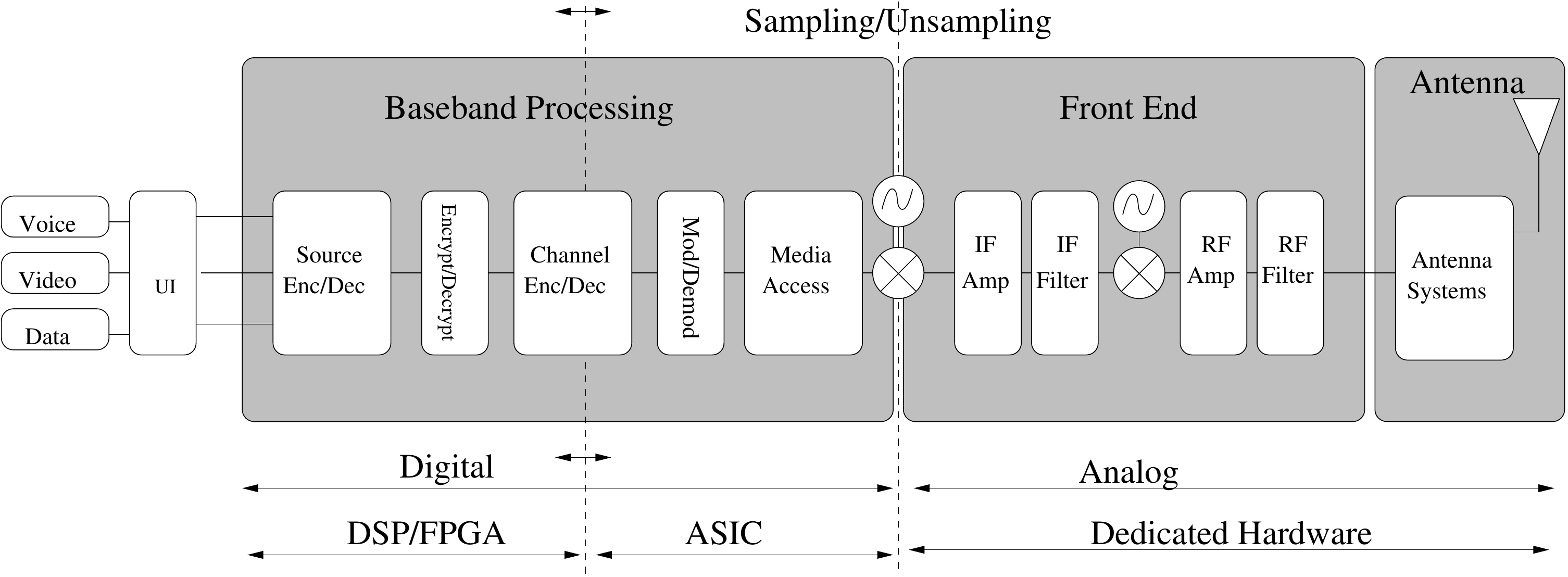}}
	\caption{Radio Block Diagram, highlighting separation between digital and analog parts, as well as programmable, configurable and fixed hardware parts.}
	\label{fig:radio}
\end{figure}

The different components of a radio system are illustrated in
Fig.~\ref{fig:radio}. Of course, all of the digital components may not
be programmable, but the bigger the programmable part ({\dsp}/{\fpga}
part on Fig.~\ref{fig:radio}), the more {\em software} the radio.
Dedicated circuits are usually needed, for which the term configurable is more adapted than programmable. 
In a typical {\sdr}, the analog part is limited to
a frequency translation down to an intermediate band which is sampled,
and all the signal processing is done digitally.  To encourage a
common meaning for the term ``{\sdr}'', the {\sdr} Forum (recently
renamed Wireless Innovation Forum) proposes to distinguish five
tiers. Tier 0 corresponds to hardware radio, Tier 1 corresponds to
software controlled radio (only the control functions are implemented
in software) and Tier 2 corresponds to software-defined Radio and is
the most popular definition of {\sdr}: the radio includes
software control of modulation, bandwidth, frequency range and
frequency bands. Tier 3 and 4 are not realistic today.

Building an \sdr terminal includes choosing a computing platform for
the digital part, a sampling frequency and a radio front-end. In
addition to the careful choice of a computing platform, the designer
must make a trade-of between sampling frequency and terminal
complexity. For instance, sampling a signal at 4.9~GHz (hence with a
10~GHz sample rate) is today not available with reasonable power
consumption. Even with an evolution to lower power \adc, a high
bandwidth \adc would produce more samples, hence require a more
powerful or specialized platform.  In this paper, we focus on the
digital part represented on the left side of Fig.~\ref{fig:radio}.


Finally, {\em Cognitive Radio} is a wireless 
system that
can sense the air, and decide to configure itself in a given mode.
Tier 2 {\sdr} platforms are natural candidates for cognitive radio
implementation but cognitive radios do not have to be \sdr.

The hardware platforms we review in the following are considered from a \sdr point of view. They target the implementation of wireless communication protocol stacks from application down to physical layer (including baseband processing and intermediate frequency conversion), for emission (\textsc{tx}) and/or reception (\textsc{rx}).

\section{Survey of Hardware Platforms for SDR}\label{sec:platforms}

In order to classify the \sdr platforms, we need to define objective
criteria. Trying to define criteria based on used technology can be
tricky, as most platform are heterogeneous. Moreover, the technology
used may not be a relevant criterion for platform users. The user
will mainly be interested in the four following features:
programmability, flexibility, energy consumption and computing power. Choosing a computing platform for a given application is a
trade-of between these cost functions.

However, from the programmer point of view, the architecture is of
major importance because it will have a crucial impact on programming
models and tools used on the platform. We finally end up with six categories: 
\begin{itemize}
\item General-purpose \cpu approach
\item Co-processor approach
\item Processor-centric approach
\item Configurable units approach
\item Programmable blocks approach
\item Distributed approach
\end{itemize}
Each approach is described in their corresponding subsections.



\subsection{General-purpose CPU approach}\label{sub:general}

\begin{figure}[h!]
	\centerline{
	\xymatrix{RF \ar[r] & *+[F]{PU} \ar@{<-->}[r] & *+[F--]{\textit{co-processor}}}
	}
	\caption{General-purpose \cpu approach with optional co-processor}\label{fig:offline_comp}
\end{figure}
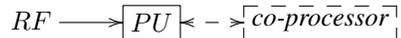

The general-purpose \cpu approach uses familiar computer processor to provide a computing platform, it is  depicted in Fig.~\ref{fig:offline_comp}. 
It offers a flexible and easy way to program the platform, but with a high energy consumption for a performance objective. 

\paragraph*{USRP}
The Universal Software Radio Peripheral (\textsc{usrp})~\cite{USRP} is representative of the General-purpose \cpu approach.
It is composed of high frequency \adc/\dac which sample the signal in intermediate frequency.
A \fpga converts and stores baseband signal.
Most of the signal processing is done by a \cpu connected to the \fpga by a \textsc{usb} link (\textsc{usrp1}) or a ethernet link (\textsc{usrp2}).
The platform is widespread and supported by third party software.
It is aimed to work with \gnu radio, but is also compatible with National Instruments LabView and Mathworks Matlab.
\paragraph*{Quicksilver}
The Quicksilver~\cite{quicksilver} module is similar in behaviour with the \textsc{usrp}.
However, it is only able to receive \textsc{rf} signals.

\paragraph*{Microsoft SORA}
Recently, Microsoft developed \textsc{sora}~\cite{Tan2011}.
This platform is connected to the computer by a PCIe bus, which permits low latency and high throughput data transmission.
It makes extensive use of modern \cpu features to perform 802.11b/g processing in real time.

With the advance of Moore Law, one could imagine that future computers will be able to compute all protocols in real time. 
However, as shown in~\cite{Ulversoy2010}, the increase in data throughput is higher than the increase in computing power.
Therefore, this kind of architecture will only be able to support past protocols, unless it can make use of higher parallelism.

\subsection{Co-processor approach}\label{sub:coproc}

In order to accelerate the signal processing, optimizations of the
General-purpose \cpu approach have been explored recently. They rely on the
addition of a co-processor to perform heavy processing. It
reduces the price to pay in terms of energy while keeping high
programmability and flexibility.

The work presented in~\cite{Horrein2011} uses a \gpu as a co-processor
in a \gnu radio flow.  It permits gains of a factor 3 to 4 in
processing speed.

\paragraph*{KUAR}
The Kansas University Agile Radio (\textsc{kuar})~\cite{Minden2007} uses an embedded \textsc{pc} associated to a \fpga.
The choice of the model of computation is left to the programmer, ranging from a full \vhdl implementation (category described in subsection~\ref{sub:FPGA}) to a full processor implementation close to the \gnu radio flow.

Other developments use generic \dsp as central processor, which provides higher efficiency while keeping high programmability.
\paragraph*{Texas Instruments}
Texas instruments offers a three-core {\dsp} with specialized symbol and chip rate accelerators.
This product provides programming flexibility for \wcdma base cells, with support for up to 64 users and different protocols~\cite{Agarwala2007}.

\paragraph*{{\imec} ADRES}
The \textsc{adres} (Architecture for Dynamically Reconfigurable Embedded Systems)~\cite{Bougard2008} developed by \imec is a coarse grain reconfigurable architecture.
It is built around a main \cpu and the \textsc{adres} accelerator.
The \textsc{adres} is seen by the processor as a \vliw, while being an array of 16 functional units (\textsc{fu}).
Each \textsc{fu} is a \simd processor, which leverages the data parallelism.
The processor is programmed using the \textsc{dresc} compiler~\cite{Mei2002}, in \ansic.
The \textsc{dresc} compiler generates code to unroll loops and compute them using the \textsc{adres} accelerator.
The \textsc{adres} is aimed at telecommunications, with benchmarks on 802.11n up to 108~Mbps and \lte up to 18~Mbps, with an average consumption of 333~mW~\cite{Bougard2008}.

\paragraph*{Hiveflex}
Hiveflex~\cite{hiveflex} produces accelerators based on many small cores.
These accelerators are scalable in number of cores, depending on the application.
All wireless protocols are targeted, from 802.11 to \lte, but no details about computing power or energy consumption are given.
The accelerators are sold as soft \textsc{ip} with HiveCC, the company \textsc{sdk}.

These architectures offer only limited task parallelism, which may reduce their efficiency. 
The next categories fill this gap using tailored architectures with heterogeneous types of processors. 

\subsection{Processor-centric approach}\label{sub:processor}

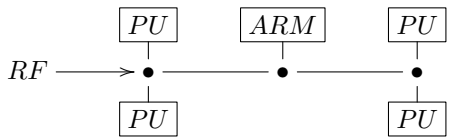
\begin{figure}[h!]
	\centerline{
	\xymatrix@R=5pt{
	& *+[F]{PU} \ar@{-}[d] & *+[F]{ARM} \ar@{-}[d] & *+[F]{PU} \ar@{-}[d] \\			
	RF \ar@{->}[r] & \bullet \ar@{-}[r] & \bullet \ar@{-}[r] & \bullet \\
	& *+[F]{PU} \ar@{-}[u] && *+[F]{PU} \ar@{-}[u]}}
	\caption{Processor-centric approach}\label{fig:processor}	
\end{figure}

One approach to get efficient and specialized platforms is to use
dedicated processors. In this approach, dedicated processors are used to
compute signal processing. The main processor, usually an {\arm},
is used for control. 

The processor-centric approach has a high programmability, but the flexibility of the platform is reduced by its specific architecture.  
 The architecture concept is depicted on
Fig.~\ref{fig:processor}.

\paragraph*{NXP EVP16}
The \textsc{nxp evp16}~\cite{Berkel2005}, presented in 2005, is composed of several units.
An \arm processor provides control and \textsc{link/mac} layers.
A conventional \dsp, a vector processor and several hardware accelerators are used for signal processing.
The vector processor is built as a vectorized pipeline and addressed as a \vliw.
It performs \textsc{umts} for a 640~kbps bandwidth at 35~MHz, with a maximum of 300~MHz~\cite{Berkel2005}.

\paragraph*{Infineon MuSIC}
Infineon built the \textsc{m}u\textsc{sic}~\cite{Ramacher2007} as a multi-\dsp solution for {\sdr}.
The control is processed by an {\arm} processor.
Signal computation is processed by 4 {\simd} {\dsp} and dedicated processors for filtering and channel encoding.
Power consumption in \wcdma mode is 382~mW for the worst case and 280~mW for normal case.
This chip is provided as a commercial solution under the name \textsc{x-gold sdr 20} by Infineon~\cite{Ramacher2011}.
It is programmed using a mix of C code and assembly code for critical processing.

\paragraph*{Sandblaster}
The Sandblaster architecture~\cite{Schulte2004} is built around 3 units, the fetch and branch, the integer and load/store and the \simd vector unit.
Task parallelism is managed by a Token Triggered Threading ($T^3$) component, which provides hardware support for multithreading.
On the SB3011~\cite{Glossner2007}, 4 sandblaster cores are integrated and controlled by an {\arm} processor.
It is programmed in \ansic with a dedicated compiler.
Maximum consumption is 171~mW for \wcdma at 384~kbps~\cite{Glossner2007}.
The SB3500 is sold as an \textsc{ip} by Optimum Semiconductor Technologies (\url{http://optimumsemi-tech.com}).

\paragraph*{University of Michigan ARDBEG}
The University of Michigan at Ann Arbor developed the \textsc{soda}~\cite{Woh2006} {\sdr} platform, and its prototype version \textsc{ardbeg}~\cite{Woh2008}.
\textsc{soda} was developed as a complete software \sdr solution.
It consists of an \arm for control and 4 \simd \dsp for signal processing.
\textsc{ardbeg} builds on that platform by adding hardware turbo decoder and optimizing \dsp for signal processing.
All programming is made using C code.
Consumption results on \textsc{ardbeg} for \wcdma and 802.11a are under 500~mW~\cite{Woh2008}.

\paragraph*{University of Dresden Tomahawk}
The University of Dresden, Germany developed the Tomahawk {\sdr} chip~\cite{Limberg2008}, aiming at \lte and \wimax.
It uses two Tensilica \risc processors for control, six vector {\dsp} and two scalar {\dsp} for signal processing, as well as \asic accelerators for filtering and decoding.
The scheduling is done by dedicated hardware and C code is used for programming.
No protocol has been implemented yet on this platform.
From the authors estimation, the platform consumption is about 1.5~W~\cite{Limberg2008}.

\subsection{Configurable units approach}\label{sub:config}

In order to offer lower energy consumption, some platforms substitute {\dsp} for configurable units.
The difference between specialized {\dsp} and configurable units is very thin, however we think that there is a frontier between these two types of devices.
 

\paragraph*{Fujitsu SDR LSI}
Fujitsu developed the \sdr \textsc{lsi}~\cite{Saito2006} in 2005.
The platform makes extensive use of hardware accelerators, associated to reconfigurable processors.
All these components are connected to a crossbar data network, and controlled by a central \arm processor.
The chip was able to run 802.11a/b with a maximum throughput of 43~Mbps~\cite{Saito2006}.

\paragraph*{{\imec} BEAR}
The \textsc{bear} \sdr platform~\cite{Palkovic2010} is the evolution of the \textsc{adres} from \imec.
It is constituted of an {\arm} processor for control and three \textsc{asip}s for coarse time synchronisation on different front ends.
Two \textsc{adres} coarse grain configurable architectures, as described in subsection~\ref{sub:coproc}, are used for baseband processing with a Viterbi accelerator.
The platform can be programmed with C or Matlab code, using the {\imec} development chain.
In terms of energy consumption, \textsc{bear} achieves 2x2 \mimo \ofdm at 108~Mbps for 231~mW~\cite{Derudder2009}.
{\imec} is licensing the \textsc{bear} platform as an \textsc{ip} block.

\paragraph*{CEA Magali}
The Magali \sdr chip~\cite{Clermidy2009} is developed by the \textsc{cea} as a telecommunication demonstration platform.
It is built on a network on chip, each peripheral having an access to the network, with an \arm processor controlling configurations.
Computation is done by coarse grain reconfigurable cores called Mephisto and reconfigurable \textsc{ip} for \ofdm, decoding and deinterleaving.
Smart memory engines are distributed on the NoC and act like \dma, while also providing data rearrangement.
The chip performs 4x2 \mimo \lte reception in the most demanding scenario with a consumption of 236~mW~\cite{Jalier2010}.

\paragraph*{EURECOM ExpressMIMO}
The Express\mimo is developed as a configurable units approach on a \fpga by \textsc{eurecom}~\cite{Nussbaum2009}. All the configurable units share a common network interface, \dma engine and microcontroller, and each as a specific configurable \textsc{ip} for data processing.
The board targets \ofdm \mimo implementation and uses  the open-source  OpenAirInterface framework~\cite{OAI}.

\paragraph*{University of Twente Annabelle}
University of Twente, Netherlands developed the Annabelle {\sdr} chip.
It is also built on a network on chip, using coarse grain reconfigurable cores.
An {\arm} processor is used for control, and accelerator modules (Viterbi, etc.) are connected to the \arm through an \textsc{amba} bus.
Only \ofdm specific benchmarks have been published at the time of submission.

\subsection{Programmable blocks approach}\label{sub:FPGA}

The last approach uses programmable blocks and is mainly constituted
of {\fpga}s. It doesn't provide programmability as it is, but great
flexibility to create tailored architectures.  Programmable blocks
offer high computing power for moderate energy consumption.

\paragraph*{XiSystem}
The XiSystem~\cite{Lodi2006} is a \vliw architecture featuring 3 concurrents datapaths, including a \picoga (Pipelined Configurable Gate Array).
The \picoga is an oriented datapath \fpga which executes specific instructions for the processor at run-time.
The development is made with C to provide code for both the \vliw and the \picoga.
It is aimed at embedded signal processing in general, with a benchmark on \textsc{mpeg2} encoding and an average consumption of 300~mW~\cite{Lodi2006}.

\paragraph*{Rice University WARP}
The Rice University has developed \textsc{warp}~\cite{warp}, an open  {\sdr} platform.
The computation is done by a Xilinx Virtex \fpga.
Programming uses \vhdl language.
An open source community is led by the Rice University to offer open source  implementations on the platform.

\paragraph*{Rutgers University WINC2R}
\textsc{winc2r} is an original platform for {\sdr} developed by the Rutgers University.
The platform is built on a \fpga, with softcore processors and accelerators.
Softcore processors can be programmed with \gnu radio.
Computation flow can be balanced on processors or accelerators, depending on the constraints.
Moreover, by using an \fpga, accelerators can be chosen and tuned during development.
802.11a has been implemented on the platform~\cite{Satarkar2009}.

\paragraph*{Lyrtech}
The Lyrtech company~\cite{lyrtech} offers development tools and platforms for {\sdr} based on \fpga.
Development is done using Simulink model-based approach.
The platform is presented as supporting \mimo \wimax.
Many other companies offer similar products based on \fpga (\cite{sundance,pentek} for instance).

\subsection{Distributed approach}\label{sub:distributed}

The distributed approach has only few elements, distributed control
representing a challenge in terms of programmability. We present here {\sdr} platforms using distributed computing.

\paragraph*{Picochip}
Picochip~\cite{Pulley2003} approaches signal processing using many small cores.
These cores are mapped on a deterministic matrix.
A C based development tool flow is provided by the company.
No benchmark is provided for this chip.
However, the company is annoucing \ofdm and 4\textsc{g} base stations as reference applications on its website.

\paragraph*{UC Davis AsAP}
The University of California at Davis developed the Asynchronous Array of Simple Processors~\cite{Truong2009} (\textsc{A}s\textsc{ap}).
This project aims at providing signal processing computation using small processors.
All processors can  communicate with their nearest neighbours, in a grid like array.
The version 2 adds hardware accelerators for \fft, Viterbi and video motion estimation, while increasing the total number of cores to 167.
Complete 802.11a/g is processed at 54~Mbps using 198~mW~\cite{Truong2009}.

\paragraph*{CEA Genepy}
\textsc{cea} Genepy~\cite{Jalier2010} is using coarser grain for its distributed approach.
It is based on Magali~\cite{Clermidy2009} technology, using the Network on Chip and the coarse grain configurable cores presented in subsection~\ref{sub:config}.
The control carried out by the \arm processor is undertaken by distributed small \risc processors.
Each cell on the network is composed of two Mephisto cores, one Smart Memory Engine and a \risc controller.
The platform is purely homogeneous, with no hardware accelerators.
In terms of computing power, 4x2 \mimo \lte reception is processed with a total consumption of 192~mW~\cite{Jalier2010}.

\section{Analysis}\label{sec:analysis}

In order to better understand each category, we summarize the main characteristics for
key-platforms that use different approaches in Table~\ref{tab:summary}. 
Energy consumption is
not defined for \fpga-based platforms because it is heavily dependent on
the configuration. Based on these key platforms, we draw conclusions on the
application fields of each category.
\begin{table}[h]
\centering
\begin{tabular}{|c|c|c|c|c|}
\hline
&	availability & application & prog. & cons. \\ \hline
USRP & commercial & N/A & C++  & $\approx$ PC \\ \hline
TI C64+ & commercial & base station & C/\textsc{asm} & 6 W \\ \hline
MuSIC & commercial & \scriptsize{WCDMA} & C/\textsc{asm} & $\leq$ 382 mW \\ \hline
Sandblaster & IP licence & \scriptsize{WCDMA} & C & 171 mW \\ \hline
ARDBEG & prototype & \scriptsize{WCDMA} & C & $\leq$ 500 mW \\ \hline
BEAR & IP licence & \scriptsize{MIMO OFDM} & {\tiny matlab/C} & 231 mW \\ \hline
Magali & prototype & \scriptsize{MIMO OFDM} & C/\textsc{asm} & 236 mW \\ \hline
ExpressMIMO & prototype &  \scriptsize{MIMO OFDM} & C & N/A \\ \hline
WARP & commercial &  \scriptsize{MIMO OFDM} & \vhdl & N/A \\ \hline
Lyrtech & commercial & N/A & {\tiny matlab/\vhdl} & N/A \\ \hline
ASAP & prototype & 802.11a/g & N/A & 198 mW \\ \hline
Genepy & prototype & \scriptsize{MIMO OFDM} & C/\textsc{asm} & 192 mW \\ \hline
\end{tabular}
\caption{Main characteristics of key platforms}\label{tab:summary}
\end{table}

If you don't want to study energy consumption nor architecture
algorithm adequacy, the general-purpose \cpu approach is the easiest
way to go. However, if you intend to study energy consumption or
computing power impact, this approach is not recommended. Indeed,
dedicated hardware platforms have very different behaviours compared
to generic processors. This makes it difficult to establish a
relationship between computing power and energy consumption for the
generic approach and others. As an example, for a given protocol,
computing requirements may vary with a factor of 100 in the
literature, depending on the architecture granularity.

In order to study computing power and have the lowest energy consumption, a 
heterogeneous approach which exploits hardware acceleration is a better starting point. In this
family, using {\dsp}s as in \imec 's solution~\cite{Palkovic2010} or configurable blocks as in Magali~\cite{Clermidy2009} is
clearly a pragmatic and efficient approach.

Unfortunately, using such a solution makes you heavily dependent on the platform architecture,
and porting a waveform to a different architecture can be tricky.
Providing a common \hal is a real challenging but promising way to develop practical multi platform \sdr.

Alternatively, the programmable blocks approach provides a flexible and efficient platform for prototyping.
It can be versatile in the architecture choice, see the radically different approaches from~\cite{warp} and~\cite{Satarkar2009} for example.

From these perspectives, we are now going to address the problem of building a cognitive \sdr testbed.

\section{Cognitive Radio Testbed}\label{sec:FIT}
\begin{figure}[htbp]
	\centerline{\includegraphics[width=\linewidth]{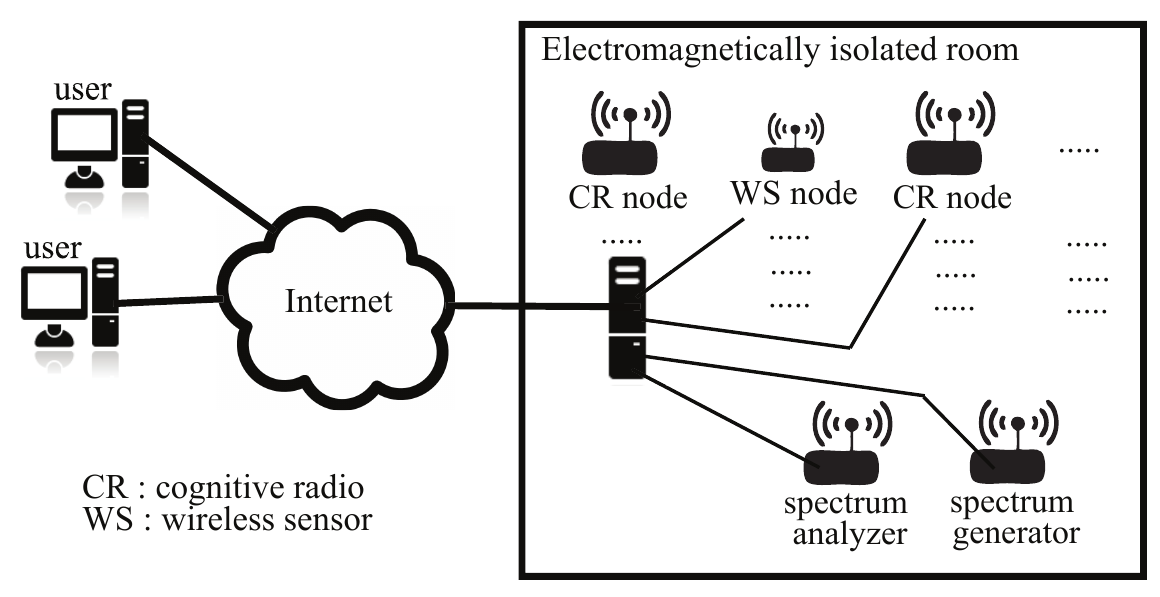}}
	\caption{Overview of the CortexLab testbed}\label{fig:fit}
\end{figure}
Before going into details on the architectural choices, we briefly review existing work on {\sdr} testbeds.
\subsection{Related work}
Large-scale cognitive radio testbeds are mandatory to develop and
evaluate the performance of upcoming \textsc{phy/mac} layers and
future cognitive radio algorithms. Whereas numerous testbeds are
available in the field of wireless communications (sensor or
802.11-oriented, see for instance Orbit developped at Winlab, Rutgers
University), only a few large-scale testbeds have been developed in
the {\sdr} and cognitive radio field. Appart from on-going projects
such as \textsc{crew}~\cite{crew} or \textsc{trial}~\cite{trial} and
some small testbeds involving less than 10 nodes, we found only one
testbed developed at Virginia Tech.,
\textsc{cornet}~\cite{Gonzalez2009}, where 48 \textsc{usrp2} with
custom RF front-ends have been dispatched in the ceilings of a
building, spanning 4 floors. The 
registered users can remotely program and run experiments on the
\textsc{usrp}s. Nodes can be programmed using the \textsc{ossie}
framework~\cite{Gonzalez2009} also developed at Virginia Tech.
\begin{figure}[htbp]
	\centerline{\includegraphics[width=\linewidth]{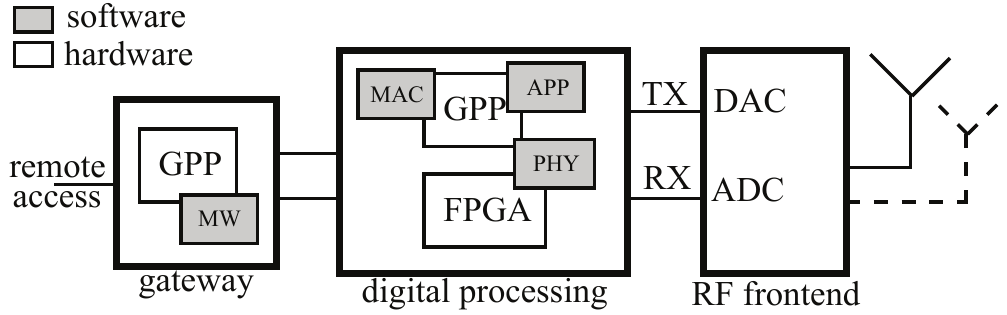}}
	\caption{Hardware/software requirements for the \cortexlab cognitive radio node.}\label{fig:fit_node}
\end{figure}
\subsection{Cortexlab}
We are currently building a new cognitive radio testbed in Lyon named
\cortexlab~\cite{CortexLab}, as part of the Future
Internet of Things~\cite{FIT} french funding, we will deploy about 50
cognitive radio nodes together with 50 wireless sensor nodes in a
electromagnetically isolated room so as to bring radio propagation
under control. The testbed will be open to the scientific community
within two years and will allow academics and industrials to conduct
real-life cognitive radio experiments. Nodes will be remotely
programmable just as if users had them on their desk. Our approach
differs from \textsc{cornet} in that 1) the topology and the room have
been selected to target reproducibility and control over the radio
propagation and 2) the nodes will have the computing power to run
\wifi/\lte in real-time at standard rates and using 2x2 \mimo. We
believe \textsc{usrp2}, even with a powerful host \textsc{pc}, cannot
achieve such a computing power. The organization of \cortexlab testbed
is illustrated on Fig.~\ref{fig:fit}.

Our main objective is to enable users to run real-time communications
with custom \textsc{app} (application such as traffic generation),
\textsc{mac} (medium access control) and \textsc{phy} layers
implementing state of the art (\wifi, Zigbee) and upcoming (\lte, \lte
adv.) standards. The programmability of the platform is a key factor
since this has to be done easily and remotely.

Following the conclusions of the previous section, we chose to mix two
types of nodes in the testbed: general-purpose \cpu nodes and
programmable blocks nodes. The general-purpose \cpu nodes should be
able to run an open source environment (\gnu radio or Open Air interface for instance) allowing rapid prototyping at slow
data rates, and the programmable ones should be able to run advanced
and \mimo \textsc{phy} layers. Fig.~\ref{fig:fit_node} shows a block
diagram of a node. The difference between the general-purpose \cpu and
the programmable node is the size of the \fpga and the function of the
\textsc{phy} layer that are assigned to it.
Note that the programming of \fpga is much
different from software programming, involving different skills and
most of the time different people. The challenge is therefore to
abstract those pieces of hardware such that they can be derived from
higher-level specifications. We are currently investigating two
approaches in this field:  high-level synthesis and 
 System-generator coupled with Matlab.

\section{Conclusion}\label{sec:conclusion}

We have reviewed existing platforms for software-defined radio and
classified them with respect to their programmability, flexibility, energy
consumption and computing power. Although
the classification we proposed is clearly arbitrary and based on our
experience, we believe this survey gives an up-to-date global view of the
available solutions. Based on our study, we saw that some 
platforms trends are emerging. In our case, as we intend to study
computing power and energy consumption while keeping programmability,
we chose a mixed \fpga/general-purpose processor platform.

A promising research direction we are investigating at the moment is
the design of a software layer able to abstract from the different
categories we have seen in this paper.

{
\bibliographystyle{ieeetr}
\bibliography{survey}
}
\end{document}